\documentclass[preprint]{revtex4-2}

\usepackage{mathtools}
\usepackage{amsmath,bm}
\usepackage{amssymb}
\usepackage{subcaption}
\usepackage{setspace}
\usepackage{dcolumn}
\usepackage{epstopdf}
\usepackage{color}

\usepackage[hidelinks]{hyperref}

\begin{document}

\title{Amplitude-dependent edge states and discrete breathers in nonlinear modulated phononic lattices}

\author{Matheus I.N. Rosa$^{a}$, Michael J. Leamy$^{b}$ and  Massimo Ruzzene$^{a}$}
\affiliation{ $^a$ Department of Mechanical Engineering, University of Colorado Boulder, Boulder CO 80309}
\affiliation{ $^b$ School of Mechanical Engineering, Georgia Institute of Technology, Atlanta GA 30332}

\date{\today}

\begin{abstract}
We investigate the spectral properties of one-dimensional spatially modulated nonlinear phononic lattices, and their evolution as a function of amplitude. In the linear regime, the stiffness modulations define a family of periodic and quasiperiodic lattices whose bandgaps host topological edge states localized at the boundaries of finite domains. With cubic nonlinearities, we show that edge states whose eigenvalue branch remains within the gap as amplitude increases remain localized,
and therefore appear to be robust with respect to amplitude. In contrast, edge states whose corresponding branch approaches the bulk bands experience de-localization transitions. These transitions are predicted through continuation studies on the linear eigenmodes as a function of amplitude, and are confirmed by direct time domain simulations on finite lattices. Through our predictions, we also observe a series of amplitude-induced localization transitions as the bulk modes detach from the nonlinear bulk bands and become discrete breathers that are localized in one or more regions of the domain. Remarkably, the predicted transitions are independent of the size of the finite lattice, and exist for both periodic and quasiperiodic lattices. These results highlight the co-existence of topological edge states and discrete breathers in nonlinear modulated lattices. Their interplay may be exploited for amplitude-induced eigenstate transitions, for the assessment of the robustness of localized states, and as a strategy to induce discrete breathers through amplitude tuning.
\end{abstract}

\maketitle

\newpage
\section{Introduction}\label{Introduction}

The discovery of topological insulators in condensed matter physics~\cite{hasan2010colloquium} has motivated the exploration of analogues in classical matter, including electromagnetic~\cite{lu2014topological}, acoustic~\cite{zhang2018topological} and elastic waves~\cite{ma2019topological}. In this context, the study of band topology predicts the existence of localized states that are immune to defects and imperfections. In 1D lattices, dimerized systems analog to the Su-Schrieffer-Heeger (SSH) model~\cite{su1979solitons} provide a simple and widely employed platform to explore topological concepts, a physical manifestation of which is the existence of localized states at an interface separating two distinct topological phases~\cite{xiao2015geometric,yang2016topological,chaunsali2017demonstrating,pal2017edge,yin2018band,wang2020robust}. More recent studies seek to enable higher dimensional topological effects in lower dimensional systems by exploiting virtual dimensions in the parameter space~\cite{qi2008topological,kraus2016quasiperiodicity,prodan2015virtual,ozawa2016synthetic}. The most common strategy considers 1D lattices whose positions or interactions are modulated according to the Aubry-Andr\'e model~\cite{aubry1980analyticity}. The modulation produces a family of periodic and quasiperiodic lattices that form Hofstadter-like spectra~\cite{hofstadter1976energy} which highlight topological gaps and edge states that are reminiscent of the quantum Hall effect in 2D electronic lattices~\cite{hatsugai1993chern}. Hofstadter spectra and topological pumping of edge states have been explored in various photonic~\cite{kraus2012topological}, acoustic~\cite{apigo2019observation,ni2019observation,chen2021landau,cheng2020experimental,chen2020physical,
xu2020physical}, and elastic platforms~\cite{apigo2018topological,martinez2018quasiperiodic,rosa2019edge,pal2019topological,xia2020topological,
gupta2020dynamics,riva2020adiabatic,grinberg2020robust,chen2019mechanical,
riva2020adiabatic,xia2021experimental,rosa2021exploring,marti2021edge}, and have been extended to 2D and 3D modulated lattices exhibiting topological properties analog to the 4D and 6D quantum Hall effects~\cite{zilberberg2018photonic,lohse2018exploring,petrides2018six,cheng2020mapping,
rosa2021topological}.

While most studies on band topology described above are conducted on linear systems, there is growing interest in the investigation of spectral properties in the presence of nonlinearities. For example, the effects of nonlinearities on the Berry phase have been studied in~\cite{liu2010berry,tuloup2020nonlinearity,zhou2022topological}, while the amplitude dependent behavior of topological states have been addressed in~\cite{dobrykh2018nonlinear,pal2018amplitude,vila2019role,tempelman2021topological}. Prior investigations also include the study of the existence of topological states and of transitions induced by nonlinearities~\cite{hadad2016self,hadad2018self,chaunsali2019self,darabi2019tunable}, of edge~\cite{ablowitz2014linear,leykam2016edge} and gap solitons~\cite{lumer2013self,solnyshkov2017chirality,deng2018metamaterials}, and of the robustness and stability of topological states~\cite{chaunsali2021stability}. Of particular relevance to this work are studies on the amplitude-dependent behavior of nonlinear 1D dimerized elastic lattices inspired by the SSH model~\cite{pal2018amplitude,vila2019role,tempelman2021topological,chaunsali2021stability}, which have inspired the present study of Aubry-Andr\'e modulated lattices. In~\cite{pal2018amplitude}, the authors illustrate the amplitude-dependent behavior of the interface states, and the associated hardening properties of the frequency response curves. These predictions were confirmed experimentally using a lattice of magnetic spinners, for which de-localization of the edge states was observed for increasing amplitude~\cite{vila2019role}. A similar behavior was illustrated in~\cite{tempelman2021topological}, where the reduction in localization in the nonlinear regime was shown for interface modes whose eigenfrequency branch tangentially approaches the nonlinear bulk bands. The study of the stability of interface modes also confirms the findings of related work~\cite{chaunsali2021stability}. These prior studies advance the understanding regarding the ability of nonlinear lattices to support topological states, while suggesting the possible use of amplitude-driven tunability of topological states.

In this study, we report on the effects of nonlinearities on the spectral properties of elastic lattices modulated according to the Aubry-Andr\'e model. In the linear regime, a Hofstadter-like spectrum is formed as a function of the stiffness modulation wavenumber, featuring non-trivial spectral band-gaps that are characterized by non-zero Chern numbers~\cite{apigo2018topological,rosa2019edge,rosa2022dynamics}. The non-trivial topology manifests itself as edge states that are localized at the boundary of finite domains, which are present for a broad family of periodic and quasiperiodic lattices. Inspired by~\cite{pal2018amplitude,vila2019role,tempelman2021topological,chaunsali2021stability}, we perform a continuation of the linear modes into the nonlinear regime, and we observe their collective behavior for increasing amplitude levels. The results highlight a number of transitions experienced by localized states due to the presence of nonlinearities. We find that the edge states remain localized at the boundaries when their frequency stays within a gap, or experience a de-localization transition as their frequency tangentially approaches a non-linear bulk band. In addition, we note that, as amplitude increases, a number of modes detach from the collective of bulk modes and transition to discrete breathers localized in one or more locations. These transitions are found to be independent of the lattice size, suggesting a general feature of nonlinear lattices. In contrast to the linear regime, where modes inside gaps are always localized at an edge (or interface), nonlinearities produce modes that are localized in multiple regions within the lattice, and that emerge as continuations of the linear bulk modes. Hence, these results illustrate the co-existence of topological edge states and discrete breathers in nonlinear modulated lattices, opening opportunities for exploring their interplay for amplitude-induced topological and localization transitions.

This paper is organized as follows: following this introduction, section~\ref{Methodsec} introduces the non-linear modulated lattices and the employed numerical simulation methods. Next, section~\ref{Linearsec} describes the behavior of the modulated lattices in the linear regime, highlighting the existence of topological edge states localized at the boundaries of finite lattices. Section~\ref{resultsec} then provides the results concerning the non-linear regime, which includes the amplitude-dependent spectra and associated mode transitions, followed by transient time-domain numerical simulations to confirm the predicted behavior. Finally, section~\ref{Conclusec} summarizes the key findings of the study and highlights possible future research directions.

\section{One-dimensional modulated phononic lattices: equations of motion and solution methods}\label{Methodsec}
\begin{figure}[t!]
\centering
\includegraphics[width=0.8\textwidth]{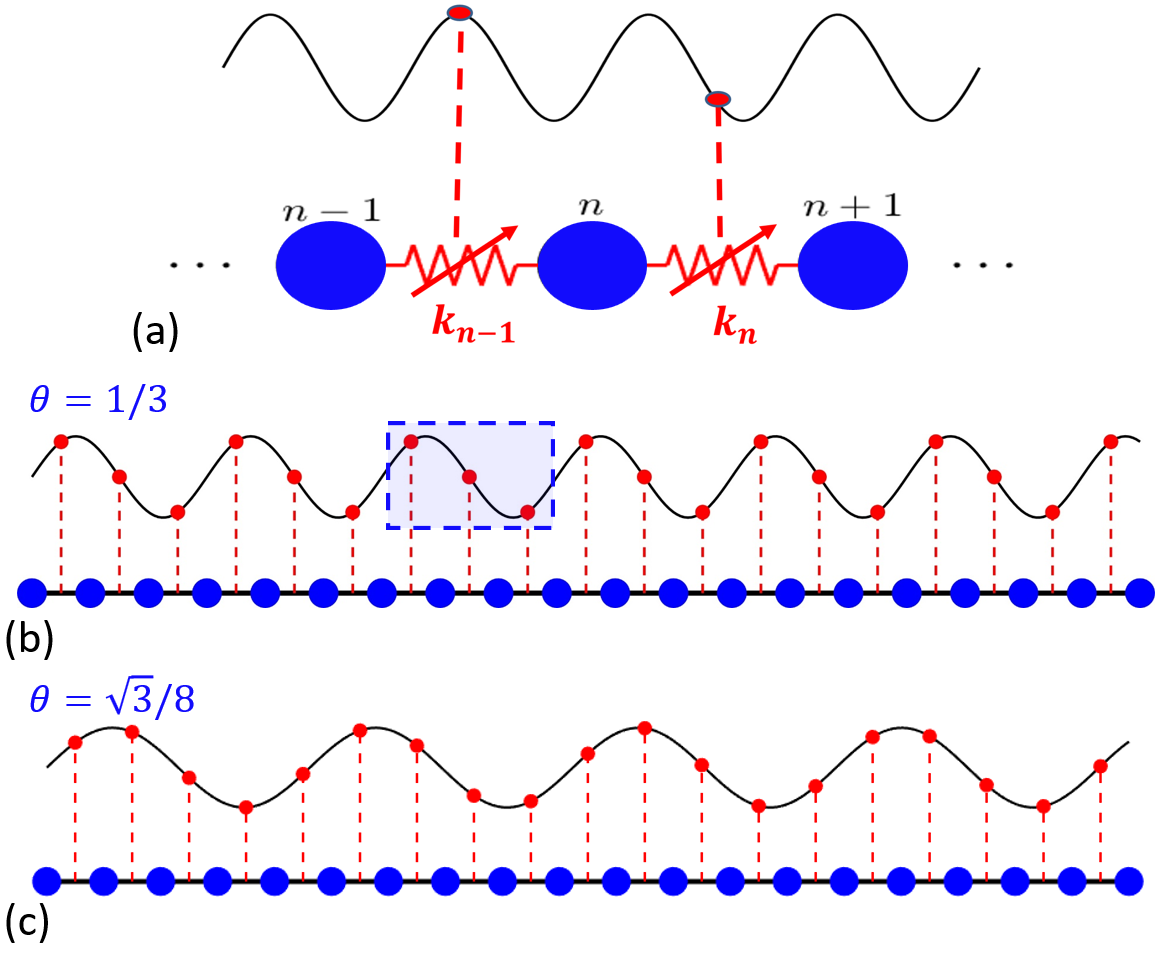}
\caption{Schematics of the nonlinear modulated elastic lattices (a). The spring constants $k_n =k_0\left[1+\lambda \cos \left( 2\pi n \theta + \phi \right) \right]$ are obtained from the sampling of a sinusoidal function~\cite{rosa2019edge}. Cubic nonlinearities with strength $\gamma$ are introduced in the expression of the spring force $f=k(\delta u + \gamma \delta u^3)$. (b) Periodic trimer lattice obtained with $\theta=1/3$, whose unit cell is highlighted by the blue box. (c) Example of quasiperiodic lattice with $\theta=\sqrt{3}/8$. }
\label{Fig1}
\end{figure}

We consider a 1D lattice of equal masses $m$, connected by springs whose stiffnesses are modulated by the sampling of a sinusoidal function (Fig.~\ref{Fig1}). In this setting, the spring constant $k_n$ connecting masses $n$ and $n+1$ is expressed as
\begin{equation}
k_n =k_0\left[1+\lambda \cos \left( 2\pi n \theta + \phi \right) \right],
\end{equation}
where $k_0$ is a stiffness offset, while $\lambda<1$ is the modulation amplitude. This modulation, inspired by the Aubry-Andr{\'e} model~\cite{aubry1980analyticity}, has been widely employed in the investigation of topological edge states in linear 1D lattices~\cite{kraus2012topological,apigo2019observation,ni2019observation,chen2021landau,cheng2020experimental,chen2020physical,
xu2020physical,apigo2018topological,martinez2018quasiperiodic,rosa2019edge,pal2019topological,xia2020topological,
gupta2020dynamics,riva2020adiabatic,grinberg2020robust,chen2019mechanical,
riva2020adiabatic,xia2021experimental,rosa2021exploring,marti2021edge}. The lattice periodicity is determined by the parameter $\theta$: rational values of the form $\theta=p/q$, where $p,q$ are co-prime integers, define periodic lattices whose unit cell comprises $q$ masses, while irrational $\theta$ values define quaisperiodic lattices with no repeating pattern of spring constants. Two examples are illustrated in Fig.~\ref{Fig1}, a periodic trimer lattice obtained with $\theta=1/3$ (b), whose unit cell comprises $3$ masses, and a quasiperiodic lattice obtained with $\theta=\sqrt{3}/8$ (c). Additionally, the modulation phase (or phason) $\phi$ is a parameter that does not affect the lattice periodicity, but produces stiffness shifts that result in the presence of edge states localized at the boundaries of finite lattices. The topological properties of this family of lattices in the linear regime have been explored in previous studies~\cite{rosa2019edge,rosa2022dynamics}, and are here investigated in the presence of cubic nonlinearities, whereby the nearest neighbor interaction is described as $f=k(\delta  + \gamma \delta ^3)$, where $\delta $ denotes the spring stretch, while $\gamma$ defines the strength of the nonlinearity. The equation of motion for mass $n$ is thus given by
\begin{align}\label{Eq1}
&m\ddot{u}_n+k_n(u_n-u_{n+1}) + k_{n-1}(u_n-u_{n-1}) \\ \nonumber
&+ \gamma k_n(u_n-u_{n+1})^3 + \gamma k_{n-1}(u_n-u_{n-1})^3 = 0,
\end{align}
with $u_n$ denoting the displacement of the $n_{th}$ mass. 

The spectral properties of finite lattices are first investigated by computing the linear eigenstates, obtained for $\gamma=0$. This is done through the solution of a standard eigenvalue problem of the form $\omega^2\mathbf{Mu}=\mathbf{Ku}$, where $\mathbf{M}$ is the mass matrix (in this case populated by a diagonal of constants $m$), and $\mathbf{K}$ is the stiffness matrix that incorporates the modulated spring constants. These linear eigenstates are then continued into the nonlinear regime for increasing amplitudes of wave motion. To that end, a harmonic balance method~\cite{vila2019role} is employed to obtain numerical estimates of the periodic solutions to the nonlinear equations of motion. These solutions are sought by assuming the motion of the $n$-th mass to be of the form 
$u_n(t)= a_n\cos(\omega t)+b_n\sin(\omega t) $, 
where $a_n, b_n$ are unknown displacement constants, while $\omega=2\pi/T$ is the assumed angular frequency of the periodic motion with period $T$, also to be determined. The method proceeds by weighting the residuals of the assumed solution against the basis provided by the considered harmonics. For a finite lattice of $N$ masses, a set of $2N$ algebraic equations with $2N+1$ unknowns is obtained by substituting the solution ansatz into Eqn.~\eqref{Eq1}, then multiplying by $\cos( \omega t)$ and, separately, by $\sin(\omega t)$, and finally integrating over the period $t \in [0, T]$. The unknowns include the $2N$ displacement constants $\bm u=\{a_1, b_1, a_2, b_2, , ..., a_N, b_N\}^T$, and the angular frequency $\omega$. An additional equation is imposed by the $\mathcal{L}_2$ norm of the displacements $||\mathbf{u}||_2=A$, where $A$ is the imposed amplitude of motion. To solve the set of algebraic equations, we start from a small amplitude $A$ and use the linear solution as a initial guess. We then progressively increase $A$ and numerically solve the algebraic equations using the previous solution as initial guess. In particular, the numerical solution follows a trust-region algorithm implemented via MATLAB's \textit{fsolve} function with default tolerance values.

For periodic lattices, the wave dispersion properties can be conveniently estimated in the linear regime by enforcing Bloch-Floquet periodicity conditions~\cite{hussein2014dynamics,rosa2019edge}. The introduction of nonlinearities causes the relationship between frequency and wavenumber to depend also on the amplitude of the wave motion. Herein, we track the amplitude-dependent nature of the dispersion bands by following a Multiple Time Scales approach, which has been successfully applied and extensively detailed in previous studies~\cite{manktelow2011multiple,fronk2017higher,fronk2019direction}.
The method is applied to first order as detailed in the appendix, and in the trimer lattice case considered in this paper ($\theta=1/3$) the following expression for the non-linear dispersion $\omega^{NL}_{j}(\mu)$ of the $j_{th}$ band is obtained:
\begin{equation}\label{EQdispNL}
\omega^{NL}_{j}(\mu)=\omega^{L}_{j}(\mu)+\frac{3}{8}\frac{\alpha_j^2}{m\omega^{L}_{j}(\mu)}c_j(\mu)\gamma,
\end{equation}
where $\mu$ is the wavenumber, $\omega^L_{j}(\mu)$ denotes the linear dispersion of the $j_{th}$ band (obtained through standard Bloch-Floquet conditions), $\alpha_j$ is the wave amplitude, and $c_j$ is a quantity defined in terms of the linear Bloch modes and stiffness values. For a given non-linearity strength $\gamma$, the equation provides the relationship between the frequency of the dispersion bands $\omega^{NL}_{j}(\mu)$ as a function of wavenumber $\mu$, which is also influenced by the wave amplitude $\alpha_j$. As it will be shown, the behavior of the finite lattice modes when continued into the nonlinear regime has a close relationship to the nonlinear dispersion bands defined above.

\section{Background: Edge states in linear modulated lattices}\label{Linearsec}
This section provides an overview of the behavior of the modulated lattices in the linear regime, focusing on the topological properties that lead to the existence of edge states. We follow a general approach that describes the behavior of both periodic and quasiperiodic lattices, which has been developed based on mathematical principles of K-theory~\cite{bellissard1992gap,prodan2016bulk}. While our description is kept brief with more focus given to the non-linear regime behavior in the next section, this approach has been applied in a series of other studies with greater level of detail. These include investigations on patterned resonators~\cite{apigo2018topological,prodan2019k}, spring-mass chains~\cite{rosa2022dynamics}, acoustic waveguides~\cite{ni2019observation,apigo2019observation}, elastic beams with stiffning or resonator inclusions~\cite{Pal_2019,xia2020topological,gupta2020dynamics,rosa2021exploring}, periodic elastic waveguides~\cite{rosa2022material} and elastic metasurfaces~\cite{pu2022topological}.

\subsection{Bulk spectra and Chern number calculation}
\begin{figure}[b!]
\centering
\includegraphics[width=0.95\textwidth]{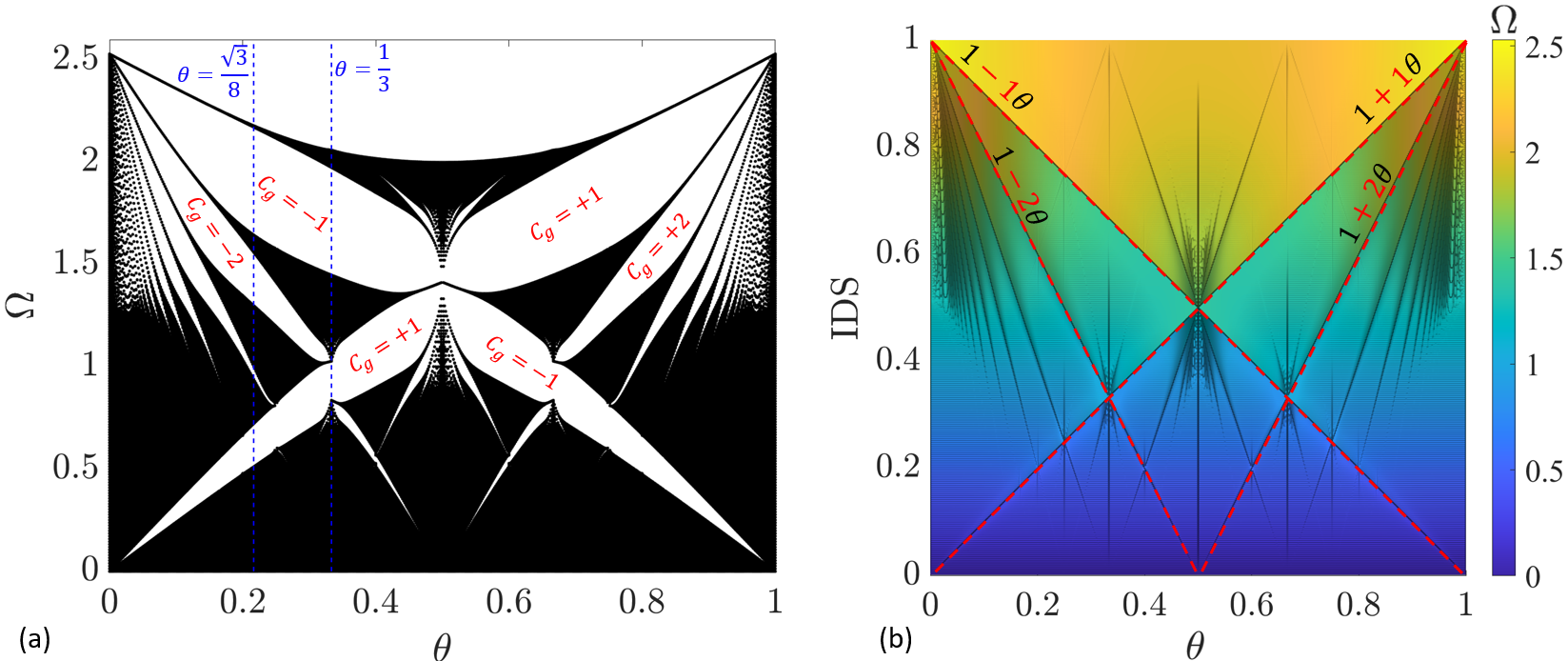}
\caption{Spectral properties of linear modulated lattices. (a) Eigenfrequencies of finite lattice with $N=600$ masses and periodic boundary conditions. Vertical dashed blue lines mark the periodic trimer lattice ($\theta=1/3$) and a quasiperiodic lattice ($\theta=\sqrt{3}/8$). (b) IDS as a function of $\theta$ with color representing frequency $\Omega$. Non-horizontal lines represent non-trivial gaps whose slope define non-zero gap labels $C_g$. Selected lines are fitted in (b) and the corresponding gaps are labeled in the spectrum shown in (a).}
\label{Fig2Linear}
\end{figure}

The properties of the modulated lattices are uncovered by analyzing its bulk spectrum, here estimated by computing the eigenfrequencies of a large finite lattice. We consider a lattice with $N=600$ masses, and compute its eigenfrequencies as a function of $\theta$ in the presence of periodic boundary conditions, i.e. by connecting the first mass to the last mass. Throughout this paper, we consider $\lambda=0.6$, and we introduce a normalized frequency $\Omega=\omega/\omega_0$, with $\omega_0=\sqrt{k_0/m}$. The results are displayed in Fig.~\ref{Fig2Linear}(a), where the frequencies are plotted as black dots and form a spectrum that resembles the Hofstadter butterfly encountered for 2D electronic lattices subject to an external magnetic field~\cite{hofstadter1976energy}. Here, the $\theta$ parameter serves as an additional dimension defining a family of periodic and quasiperiodic lattices, replacing the magnitude of the magnetic field. In the figure, band gaps are identified as the white areas featuring an absence of eigenmodes. To ensure that no modes are found inside the gaps, only commensurate $\theta$ values of the form $\theta_n = n/N$ with $ n=[0,1, ..., N]$ are considered in the computation. This choice results in perfectly periodic ring-like lattices comprising an integer number of unit cells, with no boundaries or defects to generate in-gap modes. Therefore, their frequencies sample the underlying Bloch dispersion bands for the corresponding $\theta$ value~\cite{pal2019topological}, providing a good representation of the bulk spectrum of infinite lattices. While only periodic values are considered, the spectrum is defined by continuity within the entire $[0,1]$ interval, including irrational $\theta$ values that define quasiperiodic lattices. Two vertical dashed blue lines in Fig.~\ref{Fig2Linear}(a) illustrate two examples that are used throghout this paper to exemplify the behavior of the considered lattices: a periodic trimer lattice defined by $\theta=1/3$, and a quasiperiodic lattice defined by $\theta=\sqrt{3}/8$. The frequency ranges defining bulk bands and band gaps for these lattices are respectively identified by the intersection of the vertical lines with the black and white regions of the spectrum. 

The topological properties of the spectrum are evaluated by computing the integrated density of states (IDS)~\cite{prodan2019k}. It is defined as:
\begin{equation}\label{IDS_omeg Defn.}
\text{IDS} (\Omega,\theta) = \lim_{N\to \infty} \dfrac{\sum_n [\omega_n\leq \Omega ]}{N}, 
\end{equation}
where $[\cdot]$ denotes the Iverson Brackets, which provide a value of $1$ whenever the argument is true. It consists on a simple computation whereas, for a given point $\theta,\Omega$ in the spectrum of Fig.~\ref{Fig2Linear}(a), the IDS is simply the number of modes below that frequency $\Omega$ divided by the total number of modes $N$. The computed IDS is displayed in Fig.~\ref{Fig2Linear}(b). In this representation, the colormap represents frequency $\Omega$ as a function of IDS and $\theta$. The rendering of the IDS highlights straight lines that mark discontinuities associated with the spectral gaps. They occur because the density of states does not change inside a bandgap, since it does not contain any modes. Hence, a sudden jump in frequency (color) occurs as the IDS changes from the last mode before the gap to the first mode after the gap. Non-horizontal IDS lines indicate non-trivial gaps, which are characterized by a nonzero topological invariant called the Chern number.~\cite{apigo2018topological,ni2019observation} According to the theory, the IDS inside a bandgap is expressed as
\begin{equation}
IDS(\theta)=n+C_g\theta,
\end{equation}
with the Chern gap label $C_g$ corresponding to the slope of the corresponding IDS line. A few selected lines are fitted and labeled in Fig.~\ref{Fig2Linear}(b), with the corresponding gaps labeled in the spectrum of Fig.~\ref{Fig2Linear}(a). Through such computation, the derived Chern numbers are assigned to each gap for all $\theta$ values that define it, including rational and irrational values.

\subsection{Topological edge states in finite lattices}
\begin{figure*}[t!]
	\captionsetup[subfigure]{labelformat=empty}
	\centering
	\begin{subfigure}{.95\textwidth}
	\centering
	\includegraphics[width=1\textwidth]{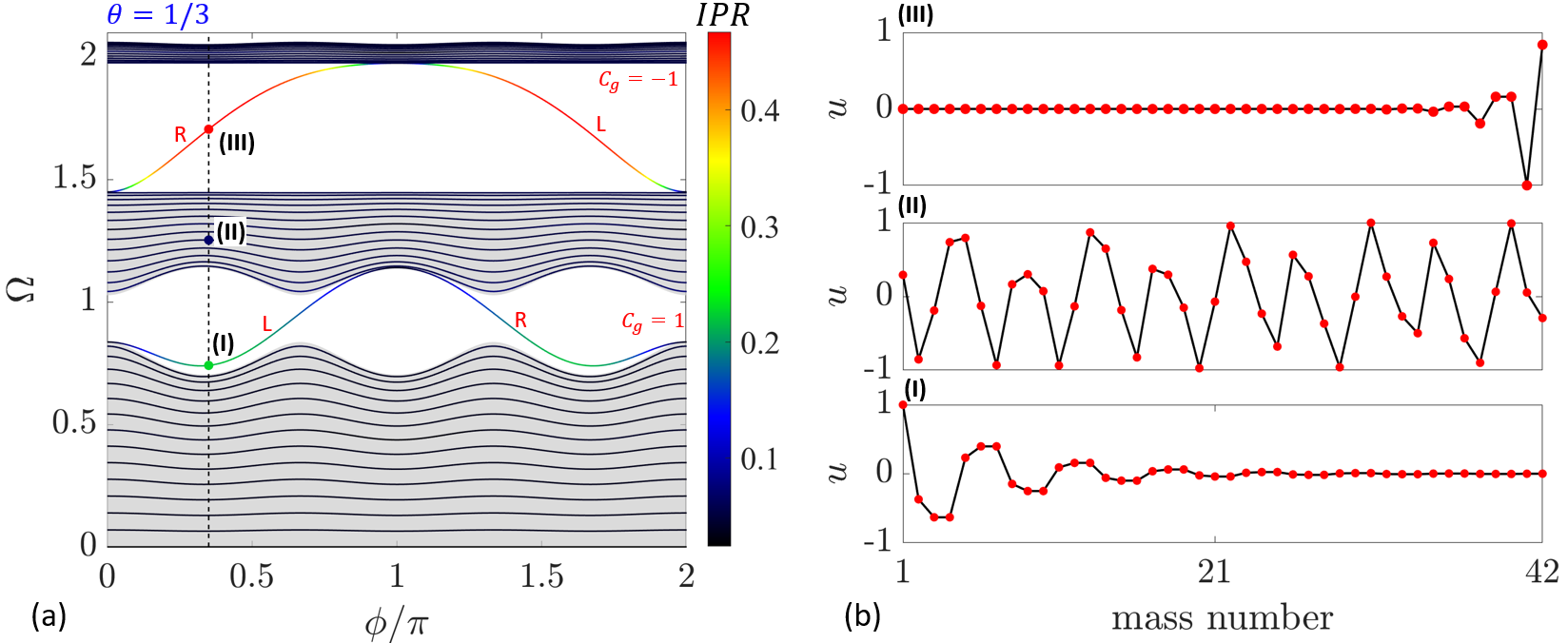}
	\end{subfigure} \\ \vspace{5mm}
	\begin{subfigure}{.95\textwidth}
	\centering
	\includegraphics[width=1\textwidth]{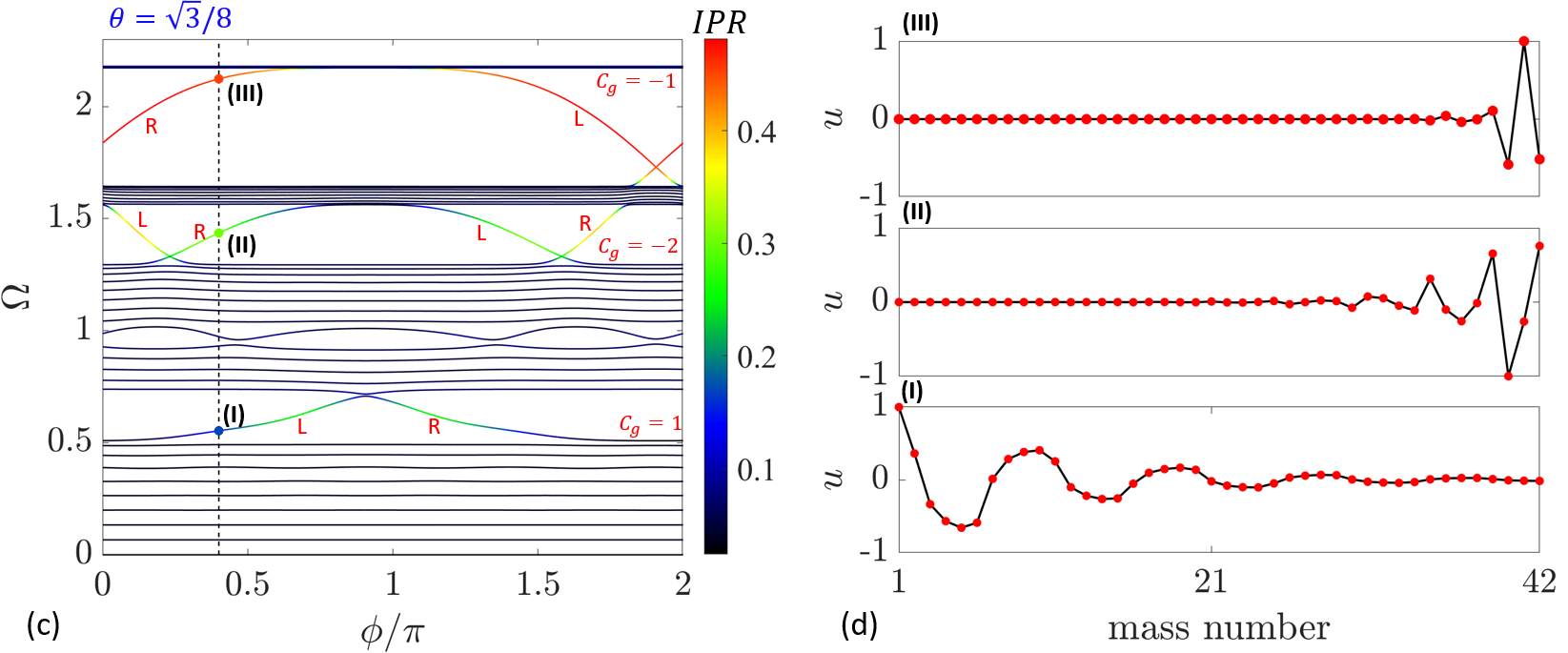}
	\end{subfigure}%
	\caption{Topological edge states in linear modulated lattices. (a,c) Eigenfrequencies as a function of stiffness modulation phase $\phi$, color-coded by the IPR, for periodic trimer lattice with $\theta=1/3$ and quasiperiodic lattice with $\theta=\sqrt{3}/8$. A finite lattice with $N=42$ masses and free-free boundary conditions is considered. The shaded gray areas in (a) denote the frequency ranges occupied by the Bloch bands for the periodic lattice. The gap labels $C_g$ are included for convenience, and the edge state branches are labeled as $R$ and $L$ for right- and left-localized, respectively. (b,d) Mode shapes for examples highlighted by the $\phi=0.35$ and $\phi=0.4\pi$ intersections (vertical dashed lines) in (a,c).  }
	\label{Fig3Linear}
\end{figure*}

The non-zero Chern numbers indicate that in-gap topological edge states will exist in finite realizations of the modulated lattices. These modes are localized at the boundaries of the lattice and define eigenvalue branches that traverse the gap as the phase $\phi$ of the stiffness modulation is varied. Such behavior is exemplified for the periodic trimer lattice with $\theta=1/3$ in Figs.~\ref{Fig3Linear}(a,b), and for the quasiperiodic lattice with $\theta=\sqrt{3}/8$ in  Figs.~\ref{Fig3Linear}(c,d), corresponding to the two cases highlighted by the blue lines in Fig.~\ref{Fig2Linear}(a). The finite lattice comprises $N=42$ masses, which in the periodic trimer lattice includes 14 unit cells. The variation of its eigenfrequencies with the phase $\phi$ under free-free boundary conditions is displayed in panels (a,c), while the mode shapes for selected eigenfrequencies are displayed in (b,d). Throughout this paper, the eigenfrequencies are color-coded according to the inverse participation ratio (IPR), which is defined as:
\begin{equation}
IPR=\frac{\sum_n u_n^4}{(\sum_n u_n^2)^2},
\end{equation}
where $u_n$ are the components of the eigenvector. High values of IPR indicate localized modes, while low values correspond to non-localized bulk modes. The plots evidence a number of mode branches within the gaps  which are localized at the boundaries, corresponding to the topological edge states. While in the linear regime described here the modes are only localized at the boundaries, the IPR is used in the following section to signal also transitions of modes that localize in other regions of the lattice. 

The features of the topological edge states are related to the corresponding gap label $C_g$. According to the theory~\cite{prodan2016bulk}, there is a total of $|C_g|$ pairs of right- and left-localized eigenvalue branches that traverse the gap as $\phi$ varies in an interval of $2\pi$. For the periodic lattice in Fig.~\ref{Fig3Linear}(a), both gaps are characterized by $|C_g|=1$ (as extracted from Fig.~\ref{Fig2Linear}(a)), and indeed feature one left-localized and one right-localized edge state. The shaded gray regions in the figure correspond to the frequencies defined by the linear dispersion bands. We note that, for periodic lattices, the Chern numbers can alternatively be extracted from the dispersion bands in the $\mu,\phi$ space as detailed in previous studies~\cite{rosa2019edge}, which produces the same results  as found here. The sign of the gap label indicates the direction at which the edge state branches migrate as they traverse the gaps. When $C_g>0$ (as in the first gap with $C_g=1$), the left-localized states migrate from the band below the gap to the band above the gap, while the right-localized states migrate in the opposite direction. The behavior is inverted for $C_g<0$, as observed in the second gap with $C_g=-1$. The modes displayed in Fig.~\ref{Fig3Linear}(b) exemplify a left-localized edge state (I), a non-localized bulk mode (II), and a right-localized edge state (III) that are defined by the $\phi=0.35\pi$ intersection (dashed vertical line) of the spectrum in (a). Similarly, the quasiperiodic lattice with $\theta=\sqrt{3}/8$ in Fig.~\ref{Fig3Linear}(c) is characterized by three main gaps with Chern gap labels of $1$, $-2$ and $-1$, as extracted from Fig.~\ref{Fig2Linear}. The number of modes traversing each gap is in agreement with the gap label, with the first and last gap exhibiting one pair of right and left-localized edge states, and the middle gap exhibiting two pairs. The modes displayed in Fig.~\ref{Fig3Linear}(d) exemplify a left-localized edge state (I), and two right-localized edge states (II,III) that are defined by the $\phi=0.4\pi$ intersection of the spectrum in (c). These two cases are selected to exemplify the general behavior of periodic and quasiperiodic lattices; other $\theta$ values would define different intersections of the spectrum of Fig.~\ref{Fig2Linear}(a), with band-gaps at different frequencies, but with the same features in terms of topological edge states and their relationship to the Chern gap labels $C_g$. In the following section, the amplitude-dependent behavior of the lattices is explored under the presence of nonlinearities. 

\section{Spectral properties of nonlinear modulated lattices}\label{resultsec}

The presence of nonlinearities introduce an amplitude dependence on the time-periodic solutions, with the linear eigenmodes detailed in the previous section defining the low-amplitude solutions. In this section we uncover the amplitude-dependent nature of such periodic solutions, showing a series of localization and de-localization transitions of the lattice modes that occur for increasing amplitude levels. We first explore the periodic trimer lattice ($\theta=1/3$), illustrating it's behavior for both positive and negative cubic nonlinearities, and also for different lattice sizes. We then characterize similar transitions that occur for the quasiperiodic lattice with $\theta=\sqrt{3}/8$. Finally, we confirm the existence of the predicted periodic solutions by simulating the free temporal evolution of the lattice motion when these are enforced as initial conditions.

\subsection{Amplitude-induced modal transitions for periodic trimer lattice ($\theta=1/3$)}

\subsubsection{Positive cubic nonlinearities}
\begin{figure*}[t!]
	\captionsetup[subfigure]{labelformat=empty}
	\centering
	\begin{subfigure}{.9\textwidth}
	\centering
	\includegraphics[width=0.5\textwidth]{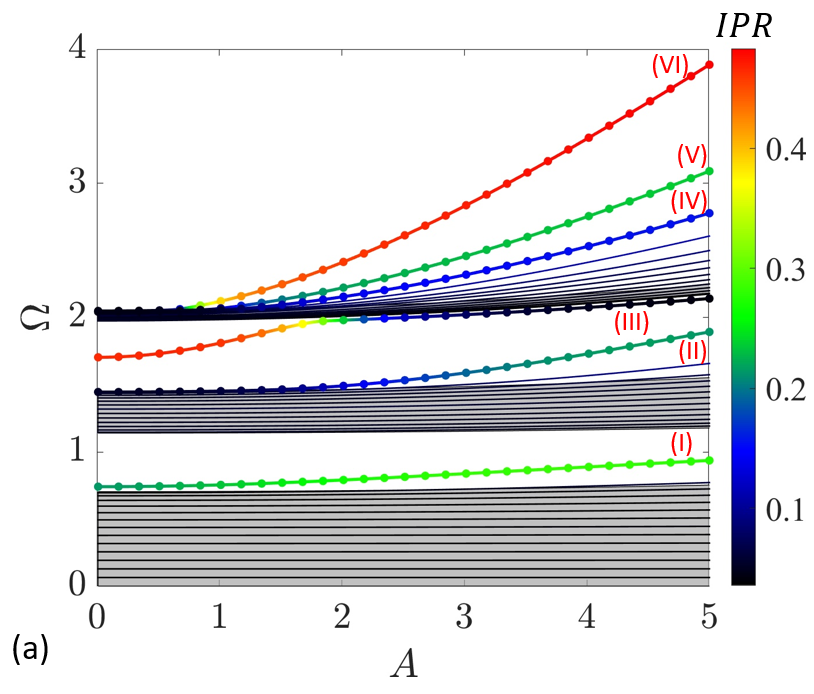}
	\end{subfigure} \\ \vspace{5mm}
	\begin{subfigure}{.95\textwidth}
	\centering
	\includegraphics[width=1\textwidth]{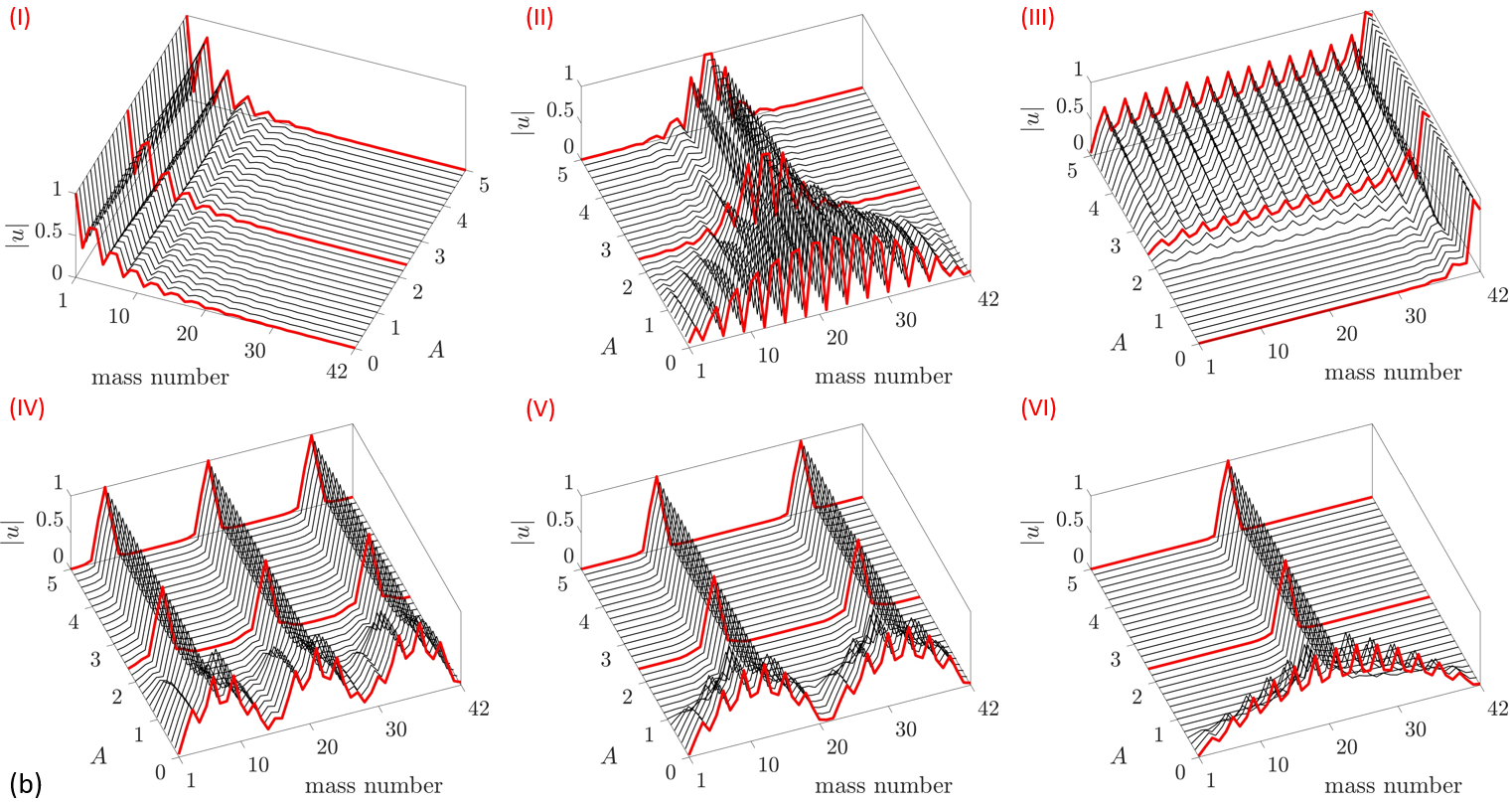}
	\end{subfigure}%
	\caption{Amplitude dependent spectrum  of periodic trimer lattice with $\theta=1/3$, $N=42$ masses, $\phi=0.35\pi$ and $\gamma=0.1$ (a). Selected highlighted branches have their mode shapes displayed in the panels of (c), for the amplitudes corresponding to the dots along the branches. The mode shapes for the initial, intermediate and final amplitude value are highlighted by red lines to enhance the visualization of their transitions.}
	\label{Fig2}
\end{figure*}

We first observe the evolution of the eigenvalue branches in terms of amplitude for the periodic trimer lattice with $\theta=1/3$ and positive cubic nonlinearities of strength $\gamma=0.1$. The results for $\phi=0.35\pi$ (vertical dashed line in Fig.~\ref{Fig3Linear}(a)) are displayed in Fig.~\ref{Fig2}(a), which were obtained as the continuation of the linear modes via the described harmonic balance approach. We observe that all eigenvalue branches experience a shift towards higher frequencies for increasing amplitudes, which is consistent with the hardening behavior of the lattice due to $\gamma>0$. The shaded gray areas correspond to the frequency ranges occupied by the nonlinear bands obtained through Eqn.~\eqref{EQdispNL} when $\mu$ is swept in the Brillouin zone $[0, \, \pi]$. These areas also experience a shift towards higher frequencies as amplitude increases. In order to match amplitudes and to conduct a comparison with the finite lattice modes, the amplitude of the Bloch wave $\alpha_j$ is linked to the amplitude $A$ of the finite lattice modes by imposing $\alpha_j=A/\sqrt{N/3}$, so that when a wave-based solution is extended to a finite lattice with $N$ masses, the resulting $\mathcal{L}_2$ norm is equal to $A$. In the linear regime, the eigenfrequencies of the finite lattice lie within the shaded regions that define the Bloch bands, with the exception of the topological modes inside the gaps that are localized at the edges. As amplitude increases, we observe that the majority of the finite lattice modes still remain concentrated in frequency regions delimited by the nonlinear dispersion bands. However, a few mode branches detach from or approach these bands, undergoing localization or de-localization transitions as evidenced by the variation of their IPR. In contrast to the linear regime, the localization induced by nonlinearities may occur in multiple regions within the lattice, not only at the edges. This behavior is illustrated for a few selected mode branches highlighted by thicker lines and dots in Fig.~\ref{Fig2}(a), whose mode shapes are displayed as a function of amplitude in Fig.~\ref{Fig2}(b). Each panel displays the variation of absolute value of the mode shape, normalized by the maximum value at each amplitude. The mode shapes for the initial, intermediate, and final amplitude value are highlighted by red lines to highlight the transitions that have occurred for these modes.

We first examine the behavior of the topological edge state branches I and III, which are the continuation of the linear modes I and III of Fig.~\ref{Fig3Linear}(a). The edge state in the first gap (mode I), which is localized at the left boundary in the linear regime, remains localized for increasing amplitudes since its branch remains within the gap. Note that the nonlinearities induce a shift of the branch to higher frequencies. However, this shift is not sufficient to cause the branch to exit the band-gap region for the considered range of amplitudes. In contrast, the branch of the edge state in the second gap (mode III), which is localized at the right boundary in the linear regime, slowly approaches the boundaries of the third nonlinear band, eventually remaining tangential to it. This causes a de-localization transition for the corresponding mode shape (Fig.~\ref{Fig2}(b)), which becomes less localized as its eigenvalue branch approaches the nonlinear band. The behavior of this second edge state is reminiscent to the de-localization transitions of topological interface states previously reported for dimerized lattices~\cite{vila2019role,tempelman2021topological}. However, it is interesting to note that the modulated lattices investigated here generally feature more than one edge state, and in the same amplitude range one edge state may remain robustly localized while the other experiences a de-localization transition.

In addition to the edge states, the modes which detach from the nonlinear bulk bands become localized in one or more regions within the lattice, as evidenced by modes II, IV, V and VI. We identify these as discrete breathers, i.e., time-periodic and spatially localized solutions. The existence of breathers has been theoretically investigated in mono-atomic lattices~\cite{willis1998breathers} and experimentally demonstrated in dimerized granular chains~\cite{boechler2010discrete}, for example. Indeed, a common factor in these prior studies is that discrete breathers emerge as continuations of the band-edge modes into the nonlinear regime for increasing amplitudes. By conducting a continuation of all the lattice modes (which we believe has not been done in prior studies), our results reveal a multitude of such breather solutions. For example, the last modes of the lattice IV-VI, which detach from the third band, become breathers localized in one, two, and three sub-regions respectively. These localization transitions seem to be triggered by the amplitude value for which the eigenvalue branch detaches from the nonlinear bulk bands. The number of regions of localization appears to be connected to the shape of the linear modes defined for $A\approx 0$: modes IV-VI are respectively characterized by three, two and one primary regions of motion in the linear regime, and then localize in the same number of sub-regions in the nonlinear regime. Of note is the transition experienced by mode II: for increasing amplitudes it detaches from the edge of the second nonlinear band and enters the region of the second gap. Through the transition, it becomes a discrete breather localized in a region near the center of the lattice. Therefore, the second gap features a right-localized edge state defined for low amplitudes (mode III), and a discrete breather at higher amplitudes (mode II). Their interplay can be potentially engineered for amplitude-induced localization transitions at a constant frequency.

\subsubsection{Negative cubic nonlinearities}
\begin{figure*}[t!]
\includegraphics[width=0.95\textwidth]{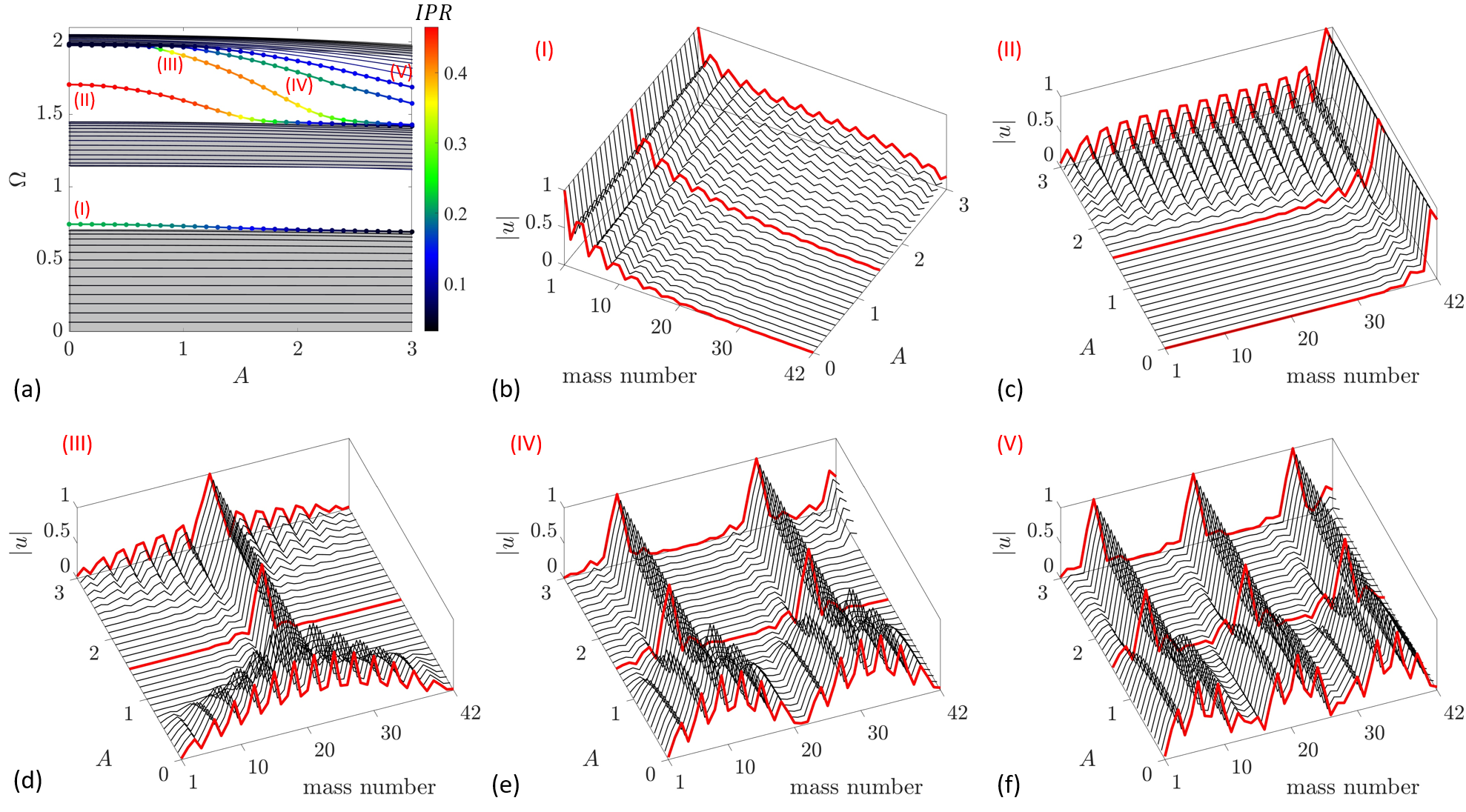}
\centering
\caption{Amplitude dependent spectrum  of periodic trimer lattice with $\theta=1/3$, $N=42$ masses, $\phi=0.35\pi$ and $\gamma=-0.1$ (a). Selected highlighted branches have their modes displayed in (b-f).}
\label{Fig3}
\end{figure*}

The spectral characteristics for negative cubic nonlinearities ($\gamma=-0.1$) are reported in Fig.~\ref{Fig3}. The results are obtained as continuation of the linear modes for the same selected phase $\phi=0.35\pi$ (vertical dashed line in Fig.~\ref{Fig3Linear}(a)). In this case, the frequencies decrease with amplitude following the typical softening behavior associated with $\gamma<0$. Both edge states (mode I and II) experience a de-localization transition as their branches tangentially approach the bulk bands below the gap. Discrete breathers localized in one (mode III), two (mode IV), and three (mode V) sub-regions emerge as the bulk modes detach from the lower edge of the third band, similar to the examples described for $\gamma>0$ above. A particularly interesting behavior is noted for mode III; it starts as an extended bulk mode for low amplitudes, and then becomes a discrete breather localized at the center of the lattice as its branch enters the region of the gap. For even higher amplitudes, its branch tangentially approaches the upper boundary of the second band, causing a reduction of localization, similar to that experienced by the edge states. Hence, this mode branch essentially migrates from one band to the other, which resembles the typical spectral flow behavior of topological states that occur upon varying a parameter such as the wavenumber, or the phase $\phi$ as shown in section~\ref{Linearsec}. Here, the spectral flow is driven by the amplitude $A$, and the behavior of the mode shape can be predicted based on whether the eigenvalue is near the bulk band (non-localized) or isolated inside the gap (localized).

\subsubsection{Influence of lattice size}

\begin{figure*}[t!]
	\captionsetup[subfigure]{labelformat=empty}
	\centering
	\begin{subfigure}{.9\textwidth}
	\centering
	\includegraphics[width=1\textwidth]{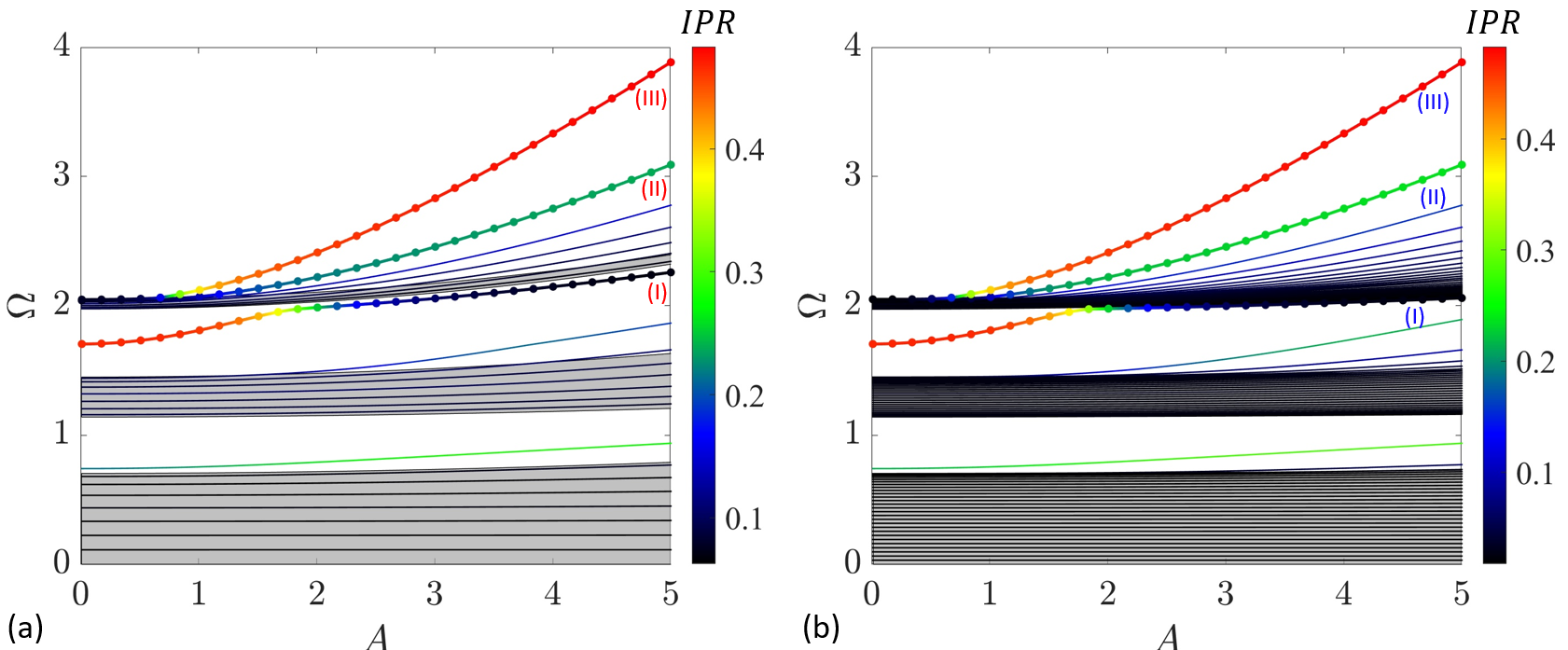}
	\end{subfigure} \\
	\begin{subfigure}{.95\textwidth}
	\centering
	\includegraphics[width=1\textwidth]{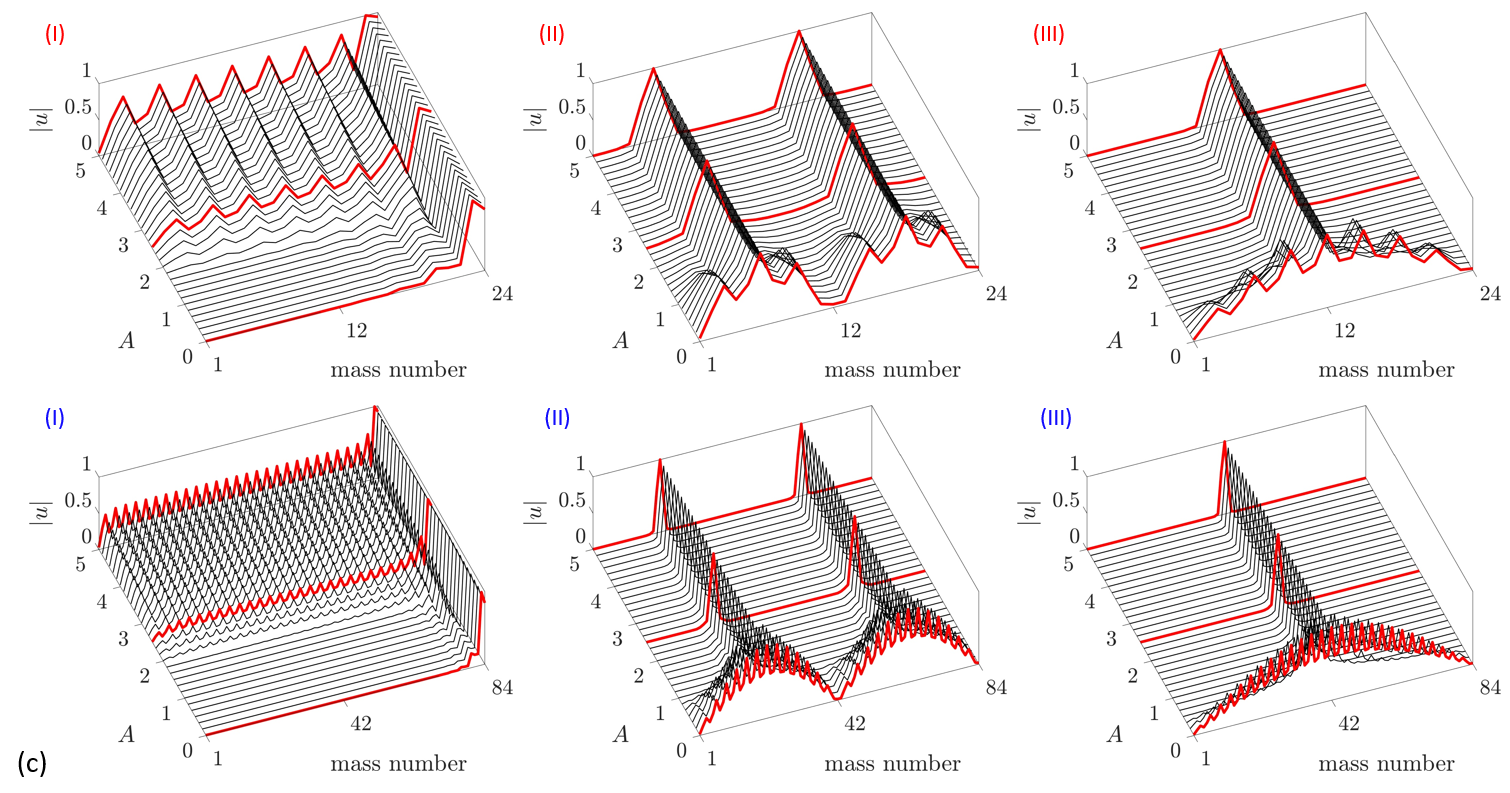}
	\end{subfigure}%
	\caption{Amplitude-dependent spectrum for trimer lattices with (a) $N=24$ and (b) $N=84$  masses, with $\theta=1/3$, $\gamma=0.1$ and $\phi=0.35\pi$. Modes highlighted in (a) and (b) are displayed as a function of amplitude in (c).}
	\label{Fig7}
\end{figure*}

While the examples described above illustrate the transitions that happen for a particular phase ($\phi=0.35\pi$), similar transitions happen for other phase values, which simply define different linear solutions as the starting point. In addition, these transitions also seem to be robust with respect to the lattice size. A few examples are illustrated in Fig.~\ref{Fig7}, which reports the amplitude-dependent spectrum for the trimer lattice ($\theta=1/3$) lattice with (a) $N=24$, and (b) $N=84$ masses for $\gamma=0.1$. Overall, both spectra are similar to that for $N=42$ shown in Fig.~\ref{Fig2}, and exhibit similar transitions. A few selected modes highlighted in Figs.~\ref{Fig7}(a,b) are displayed in Fig.~\ref{Fig7}(c), illustrating the similarity of the transitions. For instance, irrespective of lattice size, a similar de-localization transition is observed for the edge state in the second gap. Although not reported for brevity, the edge state in the first gap also remains localized as in the case of Fig.~\ref{Fig2}, since the eigenvalue branch remains within the gap. In addition, the same number of branches detach from the third bulk band and transition into discrete breathers, which are localized in the same relative regions of the lattice (center for mode III, and two regions for mode II). Our simulations also confirmed a similar behavior for other modes and lattice sizes, suggesting the potential generality of these mode transitions. 

\subsection{Amplitude-induced modal transitions for quasiperiodic lattice ($\theta=\sqrt{3}/8$)}

The results described so far illustrate the typical behavior of a periodic lattice, exemplified by the trimerized case $\theta=1/3$. It was shown that the localization or de-localization transition of the finite lattice periodic solutions respectively occur as the eigenfrequency branches detach or approach the non-linear dispersion bands. We note that other authors have defined the nonlinear bands as the frequency-amplitude (or frequency-energy) regions delimited by the first and last modes of the finite lattice~\cite{tempelman2021topological}. In the present work, this definition would include the discrete breathers IV-VI from the third band of Fig.~\ref{Fig2}, for example. Hence, our results suggest an alternative interpretation where the amplitude-dependent behavior of Bloch waves defines nonlinear bulk bands instead, since these regions correspond to plane waves propagating along the lattice with a given amplitude-frequency relation~\cite{manktelow2011multiple,fronk2017higher,fronk2019direction}. The majority of the bulk modes of a non-linear finite lattice are concentrated in such regions, and define extended modes formed by the superposition of amplitude-dependent plane waves. The discrete breathers detach from these bands and are not the superposition of traveling plane waves since they are localized. Therefore, these results suggest that the finite lattice modes do not necessarily define the nonlinear bands, and the amplitude-dependent Bloch wave solutions may provide a better representation of the non-linear bulk spectrum. 

\begin{figure}[t!]
\centering
\includegraphics[width=0.95\textwidth]{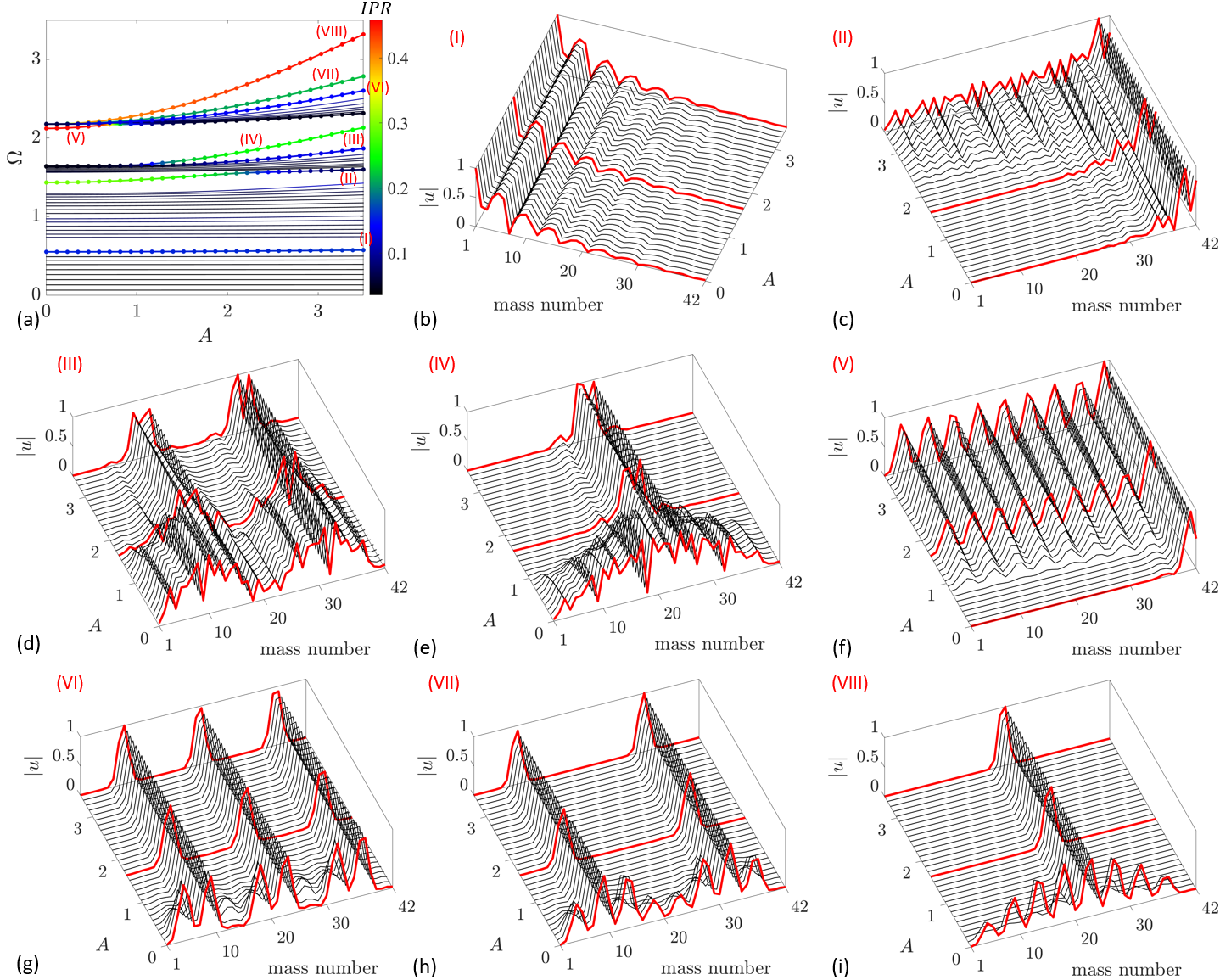}
\caption{Amplitude dependent spectrum  of quasiperiodic lattice with $\theta=\sqrt{3}/8$, $N=42$ masses, $\phi=0.4\pi$ and $\gamma=0.1$ (a). Selected highlighted branches have their modes displayed in (b-i).}
\label{Fig2QP}
\end{figure}

While the same analysis could be conducted for other periodic lattices defined by rational $\theta=p/q$, we here illustrate that such mode transitions appear more generally also for quasiperiodic lattices, that define arbitrary intersections of the spectrum in Fig.~\ref{Fig2Linear}(a) for irrational $\theta$ values. The results for a representative case $\theta=\sqrt{3}/8$ are displayed in Fig.~\ref{Fig2QP}, which show the continuation of the linear modes defined for $\phi=0.4\pi$ (vertical dashed line in Fig.~\ref{Fig3Linear}(c)) as a function of amplitude for positive cubic nonlinearities of strength $\gamma=0.1$. In the quasiperiodic case, one cannot estimate the non-linear dispersion bands due to the absence of periodicity, but the transitions of the finite lattice modes exhibit similar patterns. The linear spectrum for this example features three large band-gaps, each containing one edge state for the selected phase intersection ($\phi=0.4\pi$) as illustrated in Fig.~\ref{Fig3Linear}(d). The eigenvalue branches of the edge states all experience a shift towards higher frequencies, with the higher frequency modes experiencing a larger shift. Therefore, the edge state branch in the first gap (mode I) exhibits the smallest frequency shift and remains within the gap, while its mode shape remains localized at the left boundary (Fig.~\ref{Fig2QP}(b)). The edge state branch in the second gap (mode II) exhibits a larger frequency shift and tangentially approaches the non-linear band above the gap, which causes a de-localization transition of its right-localized mode shape (Fig.~\ref{Fig2QP}(c)). The same occurs for the edge state branch in the third gap (mode V), however the frequency shift is even larger due to its higher frequency, and the de-localization transition happens at a lower amplitude level (Fig.~\ref{Fig2QP}(f)). Similarly to the periodic case, mode branches which detach from the collective of the lattice bulk modes experience a localization transition and become discrete breathers localized in one or more regions of the lattice. These transitions are highlighted for modes (III) and (IV), which form discrete breathers localized in one and two regions (Figs.~\ref{Fig2QP}(d,e)), and for modes (VI,VII,VIII), which form discrete breathers localized in one, two and three regions (Figs.~\ref{Fig2QP}(g,h,i)).

These results highlight the features of the amplitude-induced modal transitions that generically occur for the non-linear modulated lattices, with different choices of $\theta$ and $\phi$ defining different starting points for the linear regime solutions, which may exhibit different number of bandgaps and edge states, but still feature similar modal transitions. In general, the eigenfrequencies will experience a shift towards higher or lower frequencies for positive and negative cubic nonlinearities, respectively. In the case of the edge states, they will remain localized if the frequency shift is not enough to veer its branch close to a non-linear bulk band. Otherwise, a de-localization transition occurs as its eigenvalue branch approaches the non-linear bulk band. In addition, all the linear modes at the edges of the bulk bands have a tendency to detach and transition into discrete breathers. These transitions occur faster (with respect to the amplitude) for higher frequency modes, which usually experience the larger frequency shifts. We also observe an orderly hierarchy in the nature of such localized solutions; the first mode which detaches from the bulk band becomes localized in a single region, while the second becomes localized in two sub-regions, the third in three sub-regions, and so on. This behavior was shown to be consistent across periodic lattices of different sizes and also for the quasiperiodic case presented here.

\subsection{Numerical verification of modal transitions through transient response}
\begin{figure}[b!]
\includegraphics[width=0.95\textwidth]{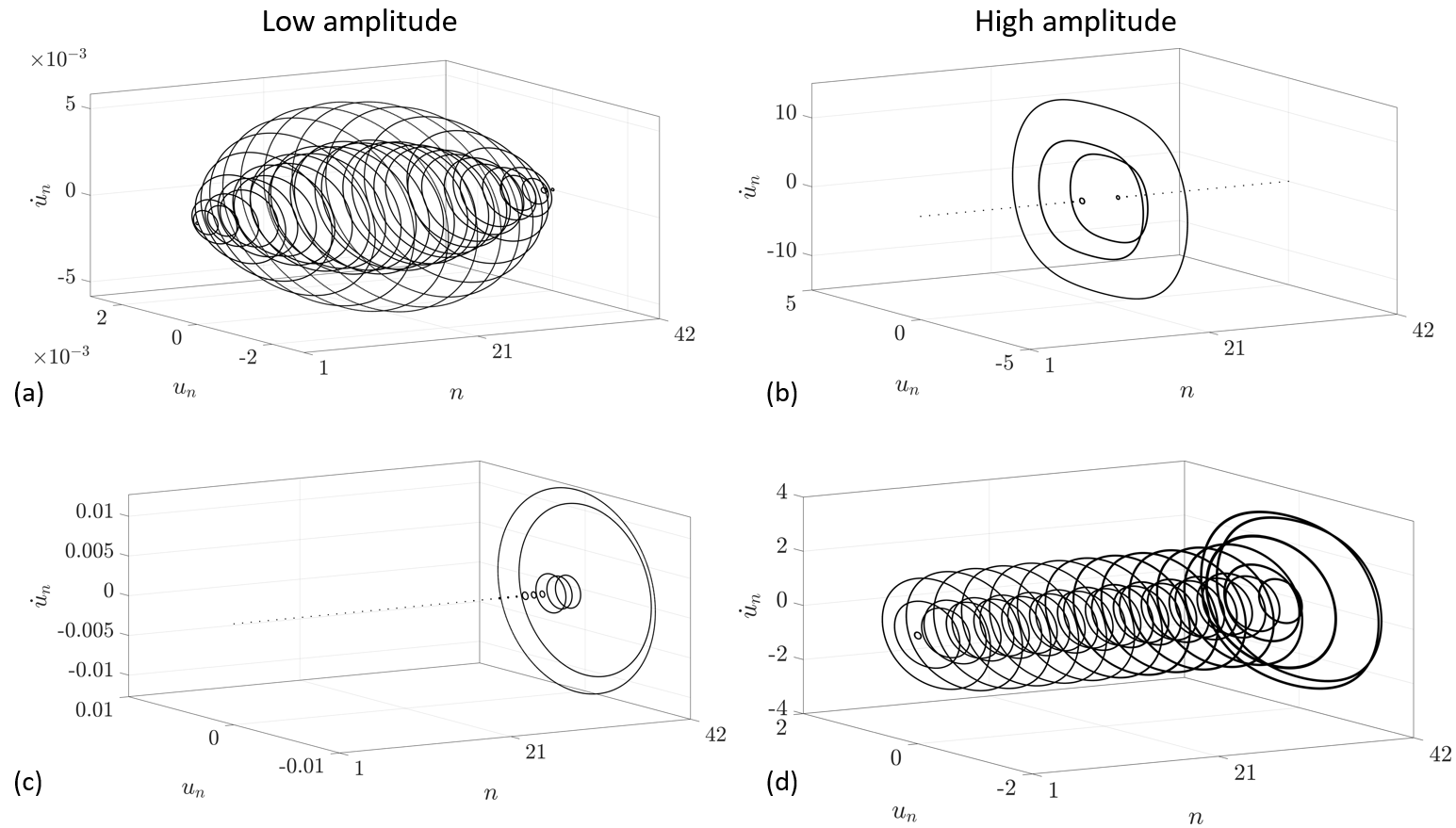}
\centering
\caption{Phase space representation plot ($u_n(t)$ vs $\dot{u}_n(t)$) illustrating the free evolution of the lattice motion characterized by periodic orbits, for different imposed initial conditions. Panels (a,b) are obtained by enforcing as initial conditions the first and last amplitudes of branch VI in Fig.~\ref{Fig2}, illustrating the localization transition to a discrete breather localized at the center of the lattice. Similarly, panels (c,d) illustrate the de-localization transition from low (c) to high (d) amplitudes experienced by the edge state (mode III of Fig.~\ref{Fig2}).}
\label{Fig4}
\end{figure}

The behavior predicted above is verified through direct time domain simulations. The existence of the modes is confirmed by first specifying the Harmonic Balance solutions as initial conditions to the finite lattice, and then simulating its free response for a total of $N_p=30$ periods of oscillation through numerical integration using Matlab's ode45 routine. A few examples are illustrated in Fig.~\ref{Fig4} using a phase space representation plot of $u_n(t)$ vs $\dot{u}_n(t)$ for each mass $n$ in the lattice. Panels (a) and (b) show results for the discrete breather identified by branch VI of Fig.~\ref{Fig2}(a), when the smallest and highest amplitudes of the branch are respectively imposed as initial conditions. The figures illustrate how the imposed solution persists and define periodic orbits for each mass in the phase space, with non-localized (a) or localized (b) character as predicted by their mode shapes, and further confirming the localization transition inducing the discrete breather localized at the center of the lattice. Similarly, panels (c) and (d) illustrate another example obtained by enforcing the first and last points belonging to branch III of Fig.~\ref{Fig2} as initial conditions. In this case, the de-localization transition experienced by the edge state from low (c) to high (d) amplitude is clearly evidenced. 

\begin{figure*}[t!]
\includegraphics[width=0.95\textwidth]{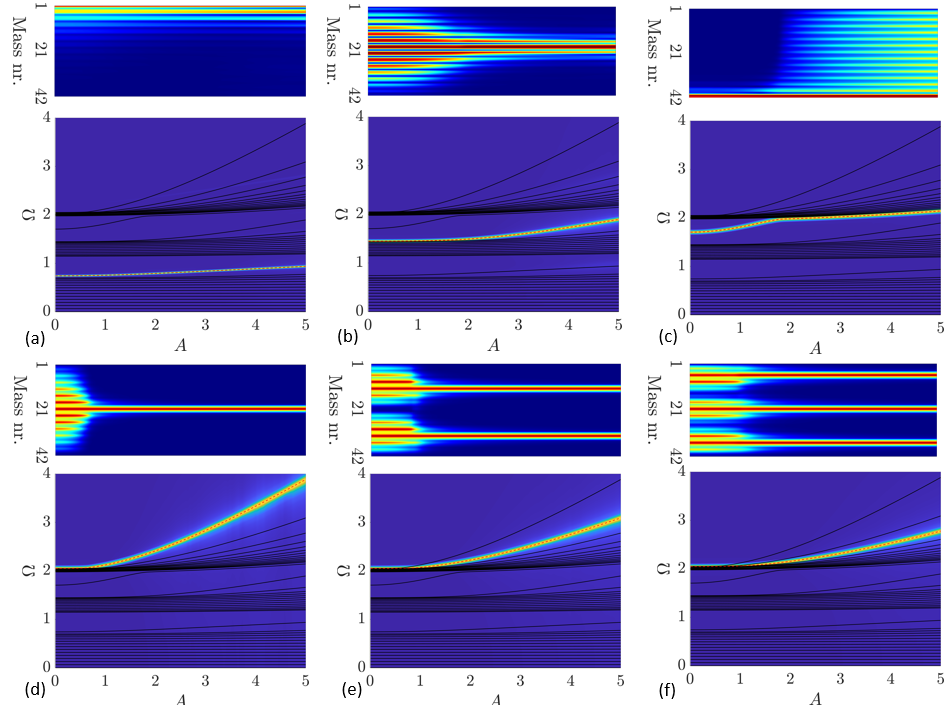}
\centering
\caption{Verification of amplitude-dependent behavior through time domain simulations for periodic trimer lattice with $\theta=1/3$, $N=42$ masses, $\phi=0.35\pi$ and $\gamma=0.1$. The bottom panels display the FT of the transient response averaged across the lattice as a function of amplitude $A$, confirming the frequency-amplitude content for each mode of interest. The top panels illustrate the RMS of the time history for each mass as a function of amplitude, confirming the predicted localization and de-localization transitions. }
\label{Fig5}
\end{figure*}

\begin{figure*}[t!]
\includegraphics[width=0.95\textwidth]{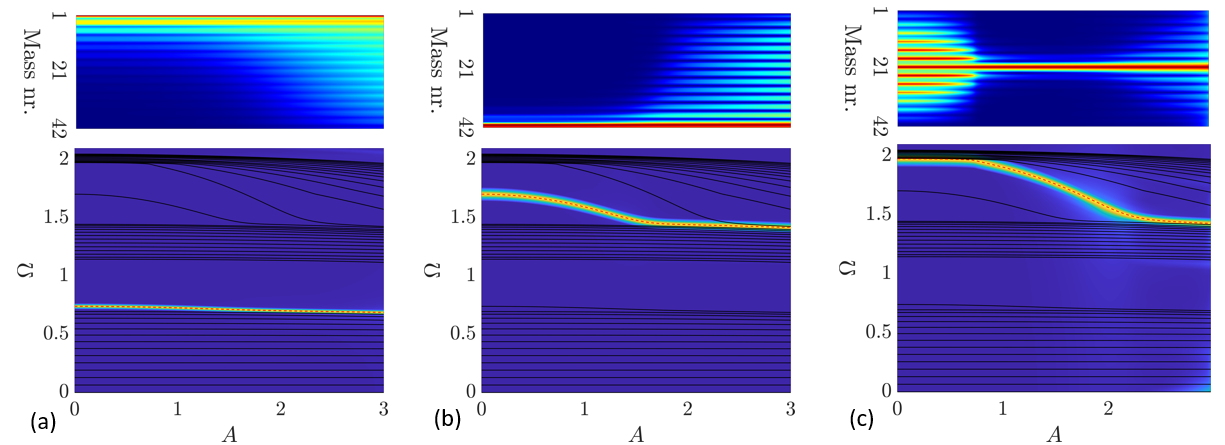}
\centering
\caption{Verification of amplitude-dependent behavior through time domain simulations for periodic trimer lattice with $\theta=1/3$, $N=42$ masses, $\phi=0.35\pi$ and $\gamma=-0.1$. The bottom panels display the FT of the transient response averaged across the lattice as a function of amplitude $A$, confirming the frequency-amplitude content for each mode of interest. The top panels illustrate the RMS of the time history for each mass as a function of amplitude, confirming the predicted localization and de-localization transitions. }
\label{Fig6}
\end{figure*}

These simulations are repeated for each branch highlighted in Fig.~\ref{Fig2}b by sweeping through the amplitude $A$, with results summarized for the 6 modes of interest in Fig.~\ref{Fig5}. Each subfigure displays the results for one mode; the bottom panel presents the Fourier Transform (FT) of the time response averaged along the entire lattice as a function of imposed amplitude $A$. On the top panel, the root mean square (RMS) of the time history for each mass is displayed as a function of amplitude. Similarly to the eigenmode plots, and for better visualization, for each individual amplitude $A$ the results are normalized to the maximum displacement along the lattice. The results confirm the amplitude-frequency content through the FTs, which exhibits good agreement with the super-imposed eigenvalue branch of the corresponding mode (dashed red lines). The RMS results in the top panels also confirm the predicted localization and de-localization transitions as a function of amplitude experienced by the modes. Another set of results for $\theta=1/3$ and $\gamma=-0.1$ is displayed for three selected modes in Fig.~\ref{Fig6} (the two edge states and the discrete breather of branch III), which also confirm the predicted transitions. Similarly, our simulations also confirmed the predicted behavior for the other cases in Figs.~\ref{Fig7} and~\ref{Fig2QP}, which are omitted for brevity. Although these results are not formal proof of stability~\cite{chaunsali2021stability,tempelman2021topological}, an important task to be carried out in future studies, they undoubtedly confirm the existence of the predicted modes as possible periodic solutions for the non-linear equations of the lattice motion. 

\section{Conclusions}\label{Conclusec}

In this paper, we investigate the amplitude-dependent behavior of nonlinear modulated phononic lattices. Our results illustrate a series of amplitude-induced localization and de-localization transitions of edge states and discrete breathers, further advancing the understanding of the nature of localized modes in nonlinear lattices. In contrast to the linear regime, where modes inside gaps are always localized at an edge (or interface), nonlinearities produce localized modes in multiple regions of the lattice, that emerge as continuations of the linear bulk modes. For periodic lattices, the mode transitions are further elucidated by including the analysis of the amplitude-dependent non-linear dispersion bands, revealing that localization or de-localization transitions respectively occur as the finite lattice modes detach or approach the regions of the non-linear bands. These amplitude-induced modal transitions are then shown to generically occur also for quasiperiodic lattices, and signal a generic feature of the family of modulated lattices. We emphasize that our results provide a framework that utilizes the linear regime solutions as a starting point, as we then observe transitions induced by increasing amplitude within the non-linear regime. Therefore, the topological properties of the edge states (i.e. Chern numbers) are still restricted to the linear regime, and a rigorous understanding or extension to the non-linear regime is still warranted, which perhaps may explain the observed de-localization transitions. Beyond the results presented herein, multiple opportunities are identified for future studies such as stability analyses~\cite{tempelman2021topological,chaunsali2021stability}, the forced response behavior, and possibilities for localization transitions between edge states and discrete breathers induced by the amplitude of motion. Additionally, the extension of these concepts to continuous non-linear elastic metamaterials~\cite{khajehtourian2014dispersion} may define a fruitful endeavor, leading towards experimental studies~\cite{zega2020experimental}.

\begin{acknowledgments}
M. I. N. R. and M. R. gratefully acknowledge the support from the National Science Foundation (NSF) through the EFRI 1741685 grant and from the Army Research office through grant W911NF-18-1-0036. M. J. L. acknowledges support from the NSF through grant 1929849.
\end{acknowledgments}

\section*{Appendix: Multiple Time Scales approach for nonlinear Bloch waves}\label{Appendix}
The dispersion of infinite lattices with nonlinear interactions is here investigated by conducting a multiple time scales analysis. The procedure described here is essentially the same as described in refs.~\cite{manktelow2011multiple,fronk2017higher,fronk2019direction}, applied to the modulated lattices. We start by writing the equations of motion for a unit cell with $q$ masses in matrix form (obtained for rational $\theta=p/q$): 
\begin{align}
&\begin{bmatrix}
m & 0 & \hdots  & 0 \\
0 & m & \hdots  & 0\\
\vdots & \vdots & \vdots & \vdots \\
0 & 0 & 0 & m
\end{bmatrix}\begin{bmatrix}
\ddot{u}_{1,j} \\ \ddot{u}_{2,j}  \\ \vdots \\ \ddot{u}_{q,j}
\end{bmatrix} + \begin{bmatrix}
k_1+k_q & -k_1 & \hdots & 0 \\
-k_1 & k_1+k_2 & -k_2 & 0 \\
\vdots & \vdots & \vdots & \vdots \\
0 & \hdots & -k_{q-1} & k_{q-1}+k_q
\end{bmatrix}\begin{bmatrix}
u_{1,j} \\ u_{2,j} \\ \vdots \\ u_{q,j}
\end{bmatrix} \nonumber \\
 &+ \begin{bmatrix}
0 & 0 & \hdots  & -k_q \\
0 & 0 & \hdots  & 0\\
\vdots & \vdots & \vdots & \vdots \\
0 & 0 & 0 & 0
\end{bmatrix}\begin{bmatrix}
u_{1,j-1} \\ u_{2,j-1} \\ \vdots \\ u_{q,j-1}
\end{bmatrix} + \begin{bmatrix}
0 & 0 & \hdots  & 0 \\
0 & 0 & \hdots  & 0\\
\vdots & \vdots & \vdots & \vdots \\
-k_q & 0 & 0 & 0
\end{bmatrix}\begin{bmatrix}
u_{1,j+1} \\ u_{2,j+1} \\ \vdots \\ u_{q,j+1}
\end{bmatrix} \nonumber \\
&+ \epsilon\begin{bmatrix}
k_1(u_{1,j}-u_{2,j})^3+k_q(u_{1,j}-u_{q,j-1})^3 \\
k_2(u_{2,j}-u_{3,j})^3+k_1(u_{2,j}-u_{1,j})^3 \\
\vdots \\
k_q(u_{q,j}-u_{1,j+1})^3+k_{q-1}(u_{q,j}-u_{q-1,j})^3
\end{bmatrix} = \begin{bmatrix}
0 \\ 0 \\ \vdots \\ 0
\end{bmatrix}
\end{align}
where $j$ denotes the index of that unit cell, and $\epsilon=\gamma$ is the strength of the cubic nonlinear interactions, now considered as a small parameter. In compact notation, we may write
\begin{equation}
\bm{M}\bm{\ddot{u}}_j+\bm{K}_{(0)}\bm{u}_j + \bm{K}_{(-1)}\bm{u}_{j-1} + \bm{K}_{(1)}\bm{u}_{j+1} + \epsilon\bm{f}^{NL}(\bm{u}_j,\bm{u}_{j-1},\bm{u}_{j+1}) = \bm{0}
\end{equation}
First, time scales of successively slower evolution are defined:
\begin{equation}
T_n=\epsilon^nt \qquad \to \qquad T_0=t, \quad T_1=\epsilon t, \quad T_2=\epsilon^2 t \quad  ...
\end{equation}
Next, a series solution for the displacements is considered
\begin{equation}
\bm{u}=\bm{u}^{(0)}_j(T_0,T_1,T_2,...) + \epsilon\bm{u}^{(1)}_j(T_0,T_1,T_2,...) + \epsilon^2\bm{u}^{(2)}_j(T_0,T_1,T_2,...) + ...
\end{equation}
Note that time derivatives can be re-written as
\begin{align}
\dot{()} &= D_0()+\epsilon D_1() + \epsilon^2 D_2() + ... \\
\ddot{()} &= D_0^2() + 2\epsilon D_0D_1() + \epsilon^2D_1^2()+2\epsilon^2D_0D_2() + ...
\end{align}
where $D_n()=\partial()/\partial T_n$. The equation of motion then becomes
\begin{align}
& D_0^2\bm{M}\bm{u}_j^{(0)} + \bm{K}_{(0)}\bm{u}^{(0)}_j + \bm{K}_{(-1)}\bm{u}^{(0)}_{j-1} + \bm{K}_{(1)}\bm{u}^{(0)}_{j+1} \nonumber \\
&+\epsilon \left( D_0^2\bm{M}\bm{u}_j^{(1)} + 2D_0D_1\bm{M}\bm{u}_j^{(0)} + \bm{K}_{(0)}\bm{u}^{(1)}_j + \bm{K}_{(-1)}\bm{u}^{(1)}_{j-1} + \bm{K}_{(1)}\bm{u}^{(1)}_{j+1} + \bm{f}^{NL} \right) +O(\epsilon^2) =0
\end{align}
We can now separate the first two ordered equations
\begin{align}
&O(\epsilon^0): \quad D_0^2\bm{M}\bm{u}_j^{(0)} + \bm{K}_{(0)}\bm{u}^{(0)}_j + \bm{K}_{(-1)}\bm{u}^{(0)}_{j-1} + \bm{K}_{(1)}\bm{u}^{(0)}_{j+1} = 0\\
&O(\epsilon^1): \quad D_0^2\bm{M}\bm{u}_j^{(1)} + \bm{K}_{(0)}\bm{u}^{(1)}_j + \bm{K}_{(-1)}\bm{u}^{(1)}_{j-1} + \bm{K}_{(1)}\bm{u}^{(1)}_{j+1} = - 2D_0D_1\bm{M}\bm{u}_j^{(0)} - \bm{f}^{NL}
\end{align}
The zeroth-order equation admits a Bloch wave solution
\begin{equation}\label{waveassume}
\bm{u}^{(0)}_{j}= \frac{A(T_1)}{2}\bm{\psi}e^{i(\mu j-\omega^L_{j}(\mu)T_0)} + \mbox{c.c}
\end{equation}
where c.c denotes complex conjugate, $A(T_1)$ is the amplitude of the Bloch mode that is only a function of $T_1$ (therefore it evolves slowly), $\bm{\psi}$ is the Bloch mode shape and $\omega^L_{j}(\mu)$ is the linear dispersion for the specified mode. Substitution of the Bloch ansatz into the zeroth-order equation yields the eigenproblem
\begin{equation}
\omega_0^2\bm{M}\bm{\psi}=\bm{K}(\mu)\bm{\psi}, \qquad \qquad \bm{K}(\mu)=\bm{K}_{(0)}+\bm{K}_{(-1)}e^{-i\mu} + \bm{K}_{(1)}e^{i\mu}
\end{equation}
whose solution gives the linear dispersion with $q$ branches and corresponding eigenvectors (note that this is the same eigenvalue problem obtained by directly enforcing Bloch conditions in the linear case). Since the $O(\epsilon^1)$ equation has the same linear kernel, we may also assume a solution in the form of
\begin{equation}
\bm{u}_j^{(1)}= \frac{B}{2}\bm{\psi}e^{i(\mu j-\omega^L_{j}(\mu)T_0)} + \mbox{c.c}
\end{equation}
In this case, the amplitude $B$ of the wave is constant since we are not carrying out $O(\epsilon^2)$ and $T_2$ terms. To identify secular terms, it is helpful to introduce modal coordinates for both the $\bm{u}^{(0)}$ and the $\bm{u}^{(1)}$ solutions, since the linear kernel of $\epsilon^0$ and $\epsilon^1$ equations can be decoupled by the Bloch modes. Hence, we define
\begin{equation}
\bm{u}_j^{(0)}=\bm{\Phi}\bm{z}_j^{(0)}e^{i\mu j} + \mbox{c.c} \qquad \mbox{,} \qquad \bm{u}_j^{(1)}=\bm{\Phi}\bm{z}_j^{(1)}e^{i\mu j} + \mbox{c.c}
\end{equation}
where $\bm{\Phi}(\mu)$ is the matrix of bloch eigenvectors, and
\begin{align}
z_{j,n}^{(0)}&=\frac{A_n(T_1)}{2}e^{-i\omega^L_{j}T_0} \\
z_{j,n}^{(1)}&=\frac{B_n}{2}e^{-i\omega^L_{j}T_0} 
\end{align}
with the relation $\omega^L_{j}(\mu)$ already defined as the dispersion obtained from solving the linear eigenvalue problem. Using this transformation, the $O(\epsilon^1)$ equation becomes
\begin{equation}
 (D_0^2\bm{M}\bm{\Phi}\bm{z}_j^{(1)} + \bm{K}(\mu)\bm{\Phi}\bm{z}_j^{(1)}+2D_0D_1\bm{M}\bm{\Phi}\bm{z}_j^{(0)} + \bm{F}^{NL} )e^{i\mu j} + \mbox{c.c} = 0
\end{equation}
while noting that the nonlinear terms may be generally written as $\bm{f}^{NL}=\bm{F}^{NL}e^{i\mu j} +\mbox{c.c}$. For non-trivial solutions, the terms multiplying $e^{i\mu j}$ must be zero, hence we get
\begin{equation}
D_0^2\bm{M}\bm{\Phi}\bm{z}_j^{(1)} + \bm{K}(\mu)\bm{\Phi}\bm{z}_j^{(1)}=-2D_0D_1\bm{M}\bm{\Phi}\bm{z}_j^{(0)} - \bm{F}^{NL}.
\end{equation}
Next, we pre-multiply by $\bm{\psi}_n^H$, yielding
\begin{equation}
 D_0^2\bm{\psi}_n^H\bm{M}\bm{\Phi}\bm{z}_j^{(1)} + \bm{\psi}_n^H\bm{K}(\mu)\bm{\Phi}\bm{z}_j^{(1)}=-2\bm{\psi}_n^HD_0D_1\bm{M}\bm{\Phi}\bm{z}_j^{(0)} - \bm{\psi}_n^H\bm{F}^{NL}.
\end{equation}
Assuming the eigenvectors are normalized as $\bm{\psi}_n^H\bm{\psi}_m=\delta_{nm}$, and that all masses on the chain have a constant mass $m$, we get that $\bm{\psi}_n^H\bm{M}\bm{\Phi}=m$ and $\bm{\psi}_n^H\bm{K}(\mu)\bm{\Phi}=(\omega^L_{n})^2m$. The updated $\epsilon^1$ equation becomes
\begin{equation}
D_0^2{z}_{j,n}^{(1)} + (\omega^L_{n})^2{z}_{j,n}^{(1)}=-2D_0D_1{z}_{j,n}^{(0)} - \frac{1}{m}\bm{\psi}_n^H\bm{F}^{NL} 
\end{equation}
To identify secular terms, we re-write the right hand side as
\begin{equation}
D_0^2{z}_{j,n}^{(1)} + (\omega^L_{n})^2{z}_{j,n}^{(1)}=\left( i\omega^L_{n}A_n'(T_1) - \frac{1}{m}\bm{\psi}_n^H\bm{F_1}^{NL} \right) e^{-i\omega^L_{n}T_0} - (\frac{1}{m}\bm{\psi}_n^H\bm{F_2}^{NL} )e^{-i3\omega^L_{n}T_0} 
\end{equation}
where the prime $()'$ denotes $D_1$ and noting that for cubic nonlinearities $\bm{F}^{NL}$ can be written as $\bm{F}^{NL}=\bm{F_1}^{NL}e^{-i\omega^L_{n}T_0}+\bm{F_2}^{NL}e^{-i3\omega^L_{n}T_0}$. Introducing a polar form for $A_n$
\begin{equation}
A_n(T_1)=\alpha_n(T_1)e^{-i\beta_n(T_1)}
\end{equation}
where $\alpha_n(T_1),\beta_n(T_1)$ are real variables, and removing secular terms gives
\begin{equation}\label{evolution1}
i\omega^L_{n}m(\alpha_n'-i\alpha_n\beta_n')e^{-i\beta_n} = \bm{\psi}_n^H\bm{F_1}^{NL} 
\end{equation}
At this point, one must specify initial conditions for the $\bm{u}_j^{(0)}$ solution that will determine the coefficients of the $\bm{F_1}^{NL} $ term. For simplicity, we assume that a single wave mode is imposed, as more complicated initial conditions would require treatment of wave-wave interactions that will not be conducted here. In particular, for the $\theta=p/q=1/3$ trimer lattice, evaluation of $\bm{\psi}_n^H\bm{F_1}^{NL}$ gives
\begin{equation}
\bm{\psi}_n^H\bm{F_1}^{NL} = \frac{3}{8}\alpha_n^3e^{-i\beta_n} c_n(\mu)  
\end{equation}
where $c_n(\mu)$ is expressed as
\begin{equation}
\begin{split}
c_n(\mu) = -2k_3\bar{\psi_1}\psi_3(|\psi_1|^2+|\psi_3|^2)e^{-i\mu} + k_3\bar{\psi_1}^2\psi_3^2e^{-2i\mu} + k_3\bar{\psi_3}^2\psi_1^2e^{2i\mu} \\
-2k_3\bar{\psi_3}\psi_1(|\psi_1|^2+|\psi_3|^2)e^{i\mu} +(k_1+k_3)|\psi_1|^4 + (k_1+k_2)|\psi_2|^4 \\
+ (4k_1|\psi_2|^2+4k_3|\psi_3|^2-2k_1(\bar{\psi_1}\psi_2 +\bar{\psi_2}\psi_1))|\psi_1|^2 + (k_2+k_3)|\psi_3|^4 \\ +(4k_2|\psi_3|^2-2k_1\bar{\psi_1}\psi_2-(2k_1\psi_1+2k_2\psi_3)\bar{\psi_2} -2k_2\psi_2\bar{\psi_3})|\psi_2|^2 \\
-2k_2(\bar{\psi_2}\psi_3+\bar{\psi_3}\psi_2)|\psi_3|^2+k_1\bar{\psi_1}^2\psi_2^2 +(k_1\psi_1^2+k_2\psi_3^2)\bar{\psi_2}^2 + k_2\bar{\psi_3}^2\psi_2^2.
\end{split}
\end{equation}
We note that through symbolic manipulation $c_n(\mu)$ is confirmed to be a purely real quantity. Considering the real and imaginary parts of Eqn.~\eqref{evolution1} yields the two evolution equations
\begin{align}
&\Re \quad \to \quad \omega^L_{n}m\beta_n'=\frac{3}{8}\alpha_n^2c_n(\mu) \\
&\Im \quad \to \quad \omega^L_{n}m\alpha_n'=0
\end{align}
The amplitude $\alpha_n$ is constant with $T_1$ and the evolution of $\beta_n$ can be obtained by simple integration 
\begin{equation}
\beta_n(T_1)=  \frac{3}{8}\frac{\alpha_n^2}{m\omega^L_{n}}c_n(\mu)T_1
\end{equation}
The $\bm{u}_j^{(0)}$ solution can now be recomposed as
\begin{equation}
\bm{u}_j^{(0)} = \alpha_n\bm{\psi_n}\cos(\mu j-\omega^{NL}_{n}(\mu)t)
\end{equation}
where $\omega^{NL}_{n}(\mu)$ is the non-linear compensated dispersion relation given by
\begin{equation}\label{fcorrected}
\omega^{NL}_{n}(\mu)=\omega^L_{n}(\mu) + \frac{3}{8}\frac{\alpha_n^2}{m\omega^L_{n}(\mu)}c_n(\mu)\epsilon
\end{equation}

\bibliographystyle{unsrt}
\bibliography{References}
\end{document}